\documentclass[lettersize,journal,hidelinks]{IEEEtran}
\usepackage{amsmath,amsfonts}
\usepackage[ruled,vlined]{algorithm2e}
\usepackage{array}
\usepackage[caption=false,font=normalsize,labelfont=sf,textfont=sf]{subfig}
\usepackage{textcomp}
\usepackage{stfloats}
\usepackage[hyphens]{url}
\usepackage{verbatim}
\usepackage{graphicx}
\usepackage{cite}

\usepackage{mathtools}
\usepackage{bbm}
\usepackage[utf8]{inputenc}
\usepackage{enumitem}
\usepackage{dutchcal}
\usepackage{adjustbox}
\usepackage{tikz}
\usetikzlibrary{shapes.geometric, arrows, positioning, shadows, decorations.pathreplacing, calc, fit}
\usepackage{multirow}
\usepackage{booktabs}
\usepackage{xspace}
\usepackage{mwe}
\usepackage{listings}
\usepackage{color}
\usepackage{orcidlink}

\usepackage{soul} 

\usepackage[skipabove=1em,framemethod=tikz]{mdframed}

\lstdefinelanguage{gitdiff}{
    basicstyle=\footnotesize\ttfamily,
    morecomment=[f][\bfseries\color{black}]{commit\ message:},
    morecomment=[f][\bfseries\color{black}]{diff\ --git},
    morecomment=[f][\color{gray}]{@@},
    morecomment=[f][\color{green!40!black}]{+\  },
    morecomment=[f][\color{green!40!black}]{+\#},
    morecomment=[f][\color{green!40!black}]{+s},
    morecomment=[f][\color{red!40!black}]{-\    },
}

\lstdefinestyle{diff}{
  escapeinside={*@}{@*},
  belowcaptionskip=1\baselineskip,
  breaklines=true,
  frame=single,
  xleftmargin=\parindent,
  language=gitdiff,
  showstringspaces=false,
  basicstyle=\scriptsize\ttfamily,
  keywordstyle=\bfseries\color{green!40!black},
  commentstyle=\itshape\color{purple!40!black},
  identifierstyle=\color{black},
  stringstyle=\color{orange},
  numbers=left,
  numbersep=5pt,
  tabsize=4,
}

\usepackage{balance}

\newcommand{\hmsc}{homoscedastic\xspace}
\newcommand{\htsc}{heteroscedastic\xspace}
\newcommand{\Hmsc}{Homoscedastic\xspace}
\newcommand{\Htsc}{Heteroscedastic\xspace}

\newcommand{\revisiontext}[1]{#1}

\newcommand{\ronerevisiontext}[1]{#1}

\newcommand{\set}[1]{\mathbb{#1}}

\usepackage{amsmath}

\begin{document}
\title{Improving Data Curation of Software Vulnerability Patches through Uncertainty Quantification}

\author{Hui~Chen, Yunhua~Zhao, Kostadin~Damevski}

\maketitle

\begin{abstract}
The changesets (or patches) that fix open source software vulnerabilities form critical datasets for various machine learning security-enhancing applications, such as automated vulnerability patching and silent fix detection. These patch datasets are derived from extensive collections of historical vulnerability fixes, maintained in databases like the Common Vulnerabilities and Exposures list and the National Vulnerability Database. However, since these databases focus on rapid notification to the security community, they contain significant inaccuracies and omissions that have a negative impact on downstream software security quality assurance tasks.

In this paper, we propose an approach employing Uncertainty Quantification (UQ) to curate datasets of publicly-available software vulnerability patches. Our methodology leverages machine learning models that incorporate UQ to differentiate between patches based on their potential utility. We begin by evaluating a number of popular UQ techniques, including Vanilla, Monte Carlo Dropout, and Model Ensemble, as well as \hmsc and \htsc models of noise. Our findings indicate that Model Ensemble and \htsc models are the best choices for vulnerability patch datasets. Based on these UQ modeling choices, we propose a heuristic that uses UQ to filter out lower quality instances and select instances with high utility value from the vulnerability dataset. Using our approach, we observe an improvement in predictive performance and a significant reduction of model training time (i.e., energy consumption) for a state-of-the-art vulnerability prediction model.

 \end{abstract}


\section{Introduction}
\label{sec:intro} 

When software vulnerabilities are reported and repaired in open-source software, the fix is recorded as a changeset (a commit or a group of semantically-related commits) in the software repository. Datasets of historic changesets (or {\em patches}) to vulnerabilities serve multiple purposes in preventing and mitigating the effects of future software vulnerabilities. They are used for automatic vulnerability patching techniques (e.g., hot-patching)\cite{altekar_opus_2005,chen_adaptive_2017}, for automatically detecting a silent fix that has not yet been publicized\cite{zhou_finding_2021,wu_enhancing_2022}, and for vulnerable code clone detection~\cite{jang_redebug_2012,kim_vuddy_2017}. Vulnerability patch datasets can also be used to extract vulnerable code to construct machine learning-based vulnerability prediction models, which have recently been showing promising results~\cite{chakraborty_deep_2022,minh_le_deepcva_2021,ghaffarian_software_2017,li_vuldeelocator_2021,li_vuldeepecker_2018}.

Large collections of vulnerability patches form the foundation for the aforementioned efforts. Government and industry have initiated concerted efforts to disseminate information about known security vulnerabilities. These include MITRE's Common Vulnerabilities and Exposures (CVE) list, the National Vulnerability Database (NVD), the Snyk Vulnerability Database, and GitHub Security Advisories. However, these efforts primarily focus on rapidly sharing details about newly found vulnerabilities to mitigate their impacts, differing from the needs of researchers who seek large datasets for building machine learning models or gathering empirical evidence to advance security-related software quality assurance~\cite{christey2013buying}. 

To close this gap, research groups have curated historic vulnerability databases using: 1) manual validation of each patch for quality; 2) automated approaches based on data selection heuristics; or 3) machine learning-based approaches trained on small manually-curated datasets. These projects face two primary challenges. First, they rely on publicly disclosed vulnerability information, such as that from the NVD, which recent research shows is often missing important information  or inaccurate~\cite{tan_locating_2021,wang_vcmatch_2022}. For example, Tan et al. found that out of 6,628 CVEs, only 66.57\% contained references to patches, and of these, 32.79\% were incorrect~\cite{tan_locating_2021}. Heuristics and machine learning models do not eliminate the errors present in these datasets and could even exacerbate them. Second, manual review of vulnerability patches to ensure high quality results in very limited dataset sizes. This effort is further complicated by the need for some familiarity with different software projects to validate each patch. Despite these costly manual efforts, the data can still suffer from significant quality issues~\cite{croft2023data}.

To address concerns about the quantity and quality of historic software vulnerability patch datasets, data curation must include the following characteristics: 1) automatic identification of patches to ensure sufficient data quantity, and 2) mitigation of the effects of inaccurately linked patches to maintain data quality for downstream uses, such as machine learning-based vulnerability prediction. 

Aside from these, prior works mostly focus on patch data quality when the patches are curated or removed from an existing dataset~\cite{croft2023data}, although the improved patch datasets are often evaluated via intended down-stream applications~\cite{tan_locating_2021,wang_vcmatch_2022}. We argue that data quality and data usefulness (i.e., the data utility value) are two related but distinct concepts and the data quality and the data usefulness exist in a spectrum. Prior works often lack of systematic or algorithmic methods to assess patch quality and usefulness. 
\ronerevisiontext{
Furthermore, existing efforts typically focus on dataset cleaning, aiming to address technical errors and inconsistencies. As we demonstrate, our data curation approach instead selects data points that maximize utility, ensuring that the resulting dataset is optimized for improving model performance and training efficiency.
}

Uncertainty Quantification (UQ) can ascertain a model's acquisition of knowledge from vulnerability patch data and control the noise and error in the vulnerability patch dataset. Building on recent research on UQ in machine learning~\cite{gal2016dropout, kendall_what_2017, ovadia_can_2019, rahaman2021uncertainty, valdenegro2022deeper}, this paper proposes a technique for integrating uncertainty quantification into software vulnerability patch data and machine learning-based models, along with a method to rank patches based on their quality and utility value. More specifically, we answer the following research questions:

\smallskip
\noindent
{\bf RQ1. What UQ techniques are capable of assessing data quality and usefulness changes in software vulnerability patch datasets?}\\
\noindent
To address this question, we compare various UQ techniques combining two data distribution modeling approaches, \hmsc and \htsc~\cite{kendall_what_2017, seitzer2022pitfalls}, with three UQ estimation methods: Vanilla, Monte Carlo Dropout, and Model Ensemble~\cite{gal2016dropout, ovadia_can_2019, rahaman2021uncertainty}. Our evaluation indicates that prediction confidence from the Vanilla approach alone is insufficient as a UQ measure. Additional uncertainty measures from Monte Carlo Dropout and Model Ensemble reveal higher uncertainty as the quality of vulnerability patches declines. Moreover, UQ models incorporating noise distributions, such as \htsc, respond better to changes in data quality and usefulness in a software vulnerability dataset. An UQ model that disentangles epistemic and aleatoric uncertainties is particularly useful. Epistemic uncertainty can serve as a proxy for the usefulness of security patches while aleatoric uncertainty the quality or the noise level of the patches.

\begin{table}[t]
\footnotesize
\centering
\caption{Vulnerability patch datasets organized according to collection approach.}
\label{tab:datasets}
\begin{tabular}{p{2.3cm}|crr|rrr}
\toprule
{\bf Dataset} & \multicolumn{3}{c|}{\textbf{Characteristics}} & \multicolumn{3}{c}{\textbf{Collection Appr.}} \\ 
& \multicolumn{1}{c}{lang.} & \multicolumn{1}{c}{\#~proj.} & \multicolumn{1}{c|}{\#~patch} &  \multicolumn{1}{c}{H1} & \multicolumn{1}{c}{H2} & \multicolumn{1}{c}{M}\\ \midrule
\textbf{SecBench}~\cite{rei_database_2017} & $>1$ & 114 & 676 & \checkmark &  & \\ 
\textbf{Lin et al.}~\cite{lin_software_2020} & C/C++ & 9 & 1471 & \checkmark & \checkmark & \\
\textbf{VulData7}~\cite{jimenez_engineering_2018} & $>1$ & 4 & 1600 & \checkmark & \checkmark & \\
\textbf{VulnCatcher}~\cite{sawadogo_learning_2020} & C/C++ & 3  & 2879 & \checkmark & \checkmark & \\ 
\textbf{SecretPatch}~\cite{wang_detecting_2019} & C/C++ & 898 & 1636 & & \checkmark & \\ 
\textbf{Big-Vul}~\cite{fan_cc_2020} & C/C++ & 348 & 10547 &  & \checkmark &  \\
\textbf{Sec. Patches}~\cite{reis_ground-truth_2021} & C/C++ & 1339 & 5942 & & \checkmark & \\
\textbf{Riom et al.}~\cite{riom_revisiting_2021} & C/C++ & 50 & 470 & & \checkmark & \\
\textbf{CVEfixes}~\cite{bhandari_cvefixes_2021} & $>1$ & 1754 & 5495 & & \checkmark & \\
\textbf{VulCurator}~\cite{nguyen_vulcurator_2022} & Python & 1  & 290 & & \checkmark & \\ 
\textbf{VulnCode-DB}~\cite{vulncode-db} & $>1$ & - & 3681 & & \checkmark & \\ 
\textbf{SAP}~\cite{ponta_manually-curated_2019} & Java & 205  & 1282 & & \checkmark & \checkmark  \\ 
\textbf{VCMatch}~\cite{wang_vcmatch_2022} & C/C++ & 10  & 1669 & & \checkmark & \checkmark  \\ 
\bottomrule
\end{tabular}

\begin{flushleft}
{\bf H1:} {\em The CVE-IDs are mentioned in 
a commit message in the project repository.}
{\bf H2:} {\em A URL to the patch is available in the NVD listing for the CVE.}
{\bf M:} {\em Patches are manually validated.}
\end{flushleft}

\end{table}

\smallskip
\noindent
{\bf RQ2. Can we use UQ to improve automatically curated vulnerability patches by selecting high quality and highly usable security patches?}\\
\noindent
We propose an algorithm to select software vulnerability patches based on epistemic and aleatoric uncertainty. The selected patches are likely to have the highest quality and highest utility values. The approach can
improve the quality of software patch datasets like those listed in Table~\ref{tab:datasets}. More importantly,
It can positively impact downstream machine learning applications, such as serving the datasets as training instances for automatic vulnerability patching techniques~\cite{altekar_opus_2005,chen_adaptive_2017}, for automatically detecting silent fixes that have not yet been publicized~\cite{zhou_finding_2021,wu_enhancing_2022}, and for vulnerable code clone detection~\cite{jang_redebug_2012,kim_vuddy_2017}. To exhibit these, we conduct an experiment that indicates that our approach can select security patches to improve the predictive performance and computational time of software vulnerability prediction, a representative downstream use case of software vulnerability patch data.

\smallskip

Answering these two RQs leads to the following contributions:
\begin{enumerate}
    \item An algorithm for automated and systematic curation of vulnerability patches based on UQ, which represents a first attempt to use both epistemic uncertainty and aleatoric uncertainty to inform the usefulness and the quality of security patches.
    \item Empirical evidence that informs the design of the UQ-based vulnerability patch curation algorithm.
    \item Statistically significant improvement of predictive performance and reduction of computational cost of a state-of-the-art software vulnerability prediction model.
\end{enumerate}
Thus, our approach: 1) has practical implications for reducing training time and improving prediction accuracy across various applications involving pre-collected vulnerability data; 2) can expand existing datasets by curating high-quality, high-utility data instances from open-source projects; and 3) encourages further, much-needed research on the application of UQ to this problem.

 \section{Aleatoric and Epistemic Uncertainty in Software Vulnerability Patch Data Curation}
\label{sec:patchdata}

\begin{figure*}[t]
\centering    \includegraphics[width=0.99\textwidth]{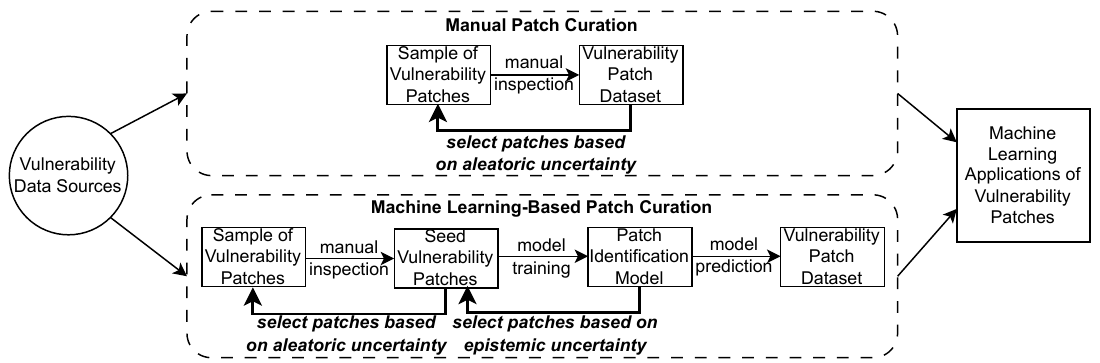}
    \caption{Sources of aleatoric and epistemic uncertainty in manual and machine learning-based vulnerability patch identification.}
    \label{fig:vulpatches}
\end{figure*}

As shown in Table~\ref{tab:datasets}, numerous efforts have been made to curate software vulnerability patch data. The primary approaches rely on heuristics like searching for CVE-IDs in commit messages and examining URLs provided in the NVD listing for CVEs (i.e., H1 and H2 in Table~\ref{tab:datasets}), which are inherently noisy and incomplete. For instance, many links gathered through H2 do not point to the actual patch but rather to related code, while CVE-IDs in commits for H1 are infrequent and can be part of tangled commits that include changes unrelated to patching the vulnerability. Ensuring the high quality of vulnerability patch datasets, therefore, requires significant manual effort. This is reflected in the relatively small size of the manually validated datasets in Table~\ref{tab:datasets}.For example, the SAP dataset~\cite{ponta_manually-curated_2019} contains 1,282 patches manually curated by teams of developers at SAP through their daily work. Such manual data curation approaches face two main challenges. Firstly, they are difficult to scale due to the extensive manual inspection required and the limited availability of software security expertise. Secondly, they cannot address the issue of missing or broken links between CVEs and patches in the NVD.

Recently, machine learning-based approaches have emerged. In ML-based approaches, researchers manually curate an initial high-quality set of ``seed'' vulnerability patches and use them to train a machine learning model, which then automatically identifies vulnerability patches~\cite{zhou_automated_2017, cabrera_commit2vec_2021, wang_patchrnn_2021, zhou_finding_2021, zhou_spi_2021, wu_enhancing_2022, tan_locating_2021, wang_vcmatch_2022}. Thus, ML-based approaches reduce the manual effort to curate larger datasets; however, they still suffer from two shortcomings: whether the instances curated are of low-quality or whether the instances contribute to improving down-stream applications, such as vulnerability prediction. Motivated by recent advances in uncertainty quantification in ML, we propose a data curation framework that addresses these two shortcomings as illustrated in Figure~\ref{fig:vulpatches}, which depicts the conceptual framework of the proposed approach.

Researchers distinguish between two types of uncertainty: aleatoric and epistemic uncertainty~\cite{kendall_what_2017}. Aleatoric uncertainty arises from the data generation and curation process, which encompasses factors such as noise in the data and the representation of data features, both of which influence training data quality~\cite{huseljic2021separation}. Conversely, epistemic uncertainty arises from inadequate modeling~\cite{hullermeier2021aleatoric}, i.e., uncertainty in the model's parameters and structures~\cite{kendall_what_2017}, indicating the model's lack of knowledge about the problem space.

Due to the coexistence of manual and ML-based approaches in vulnerability patch dataset curation, both aleatoric and epistemic uncertainties are present in patch curation~\cite{gal2016dropout, kendall_what_2017,abdar2021review}. Aleatoric uncertainty is inherently part of manual data curation, as evidenced by the pervasive errors in the NVD observed by Tan et al.~\cite{tan_locating_2021}. It also arises in ML-based vulnerability patch curation, which employs a machine learning model based on features that determine the relationship between a patch and a vulnerability (e.g., using textual similarity between the commit log and the vulnerability description). There are two primary sources of aleatoric uncertainty in this process. First, the features may not capture all relevant information concerning the origination of software vulnerability patches. Second, the curation process itself can introduce errors and noise. For example, ML-based approaches like Patchscout~\cite{tan_locating_2021} and VCMatch~\cite{wang_vcmatch_2022} identify certain feature elements, such as file locations in NVD records and function locations in committed code, using regular expression patterns manually derived from a small sample set, which can omit some feature elements and also introduce false positives, as illustrated in Figure~\ref{fig:vulpatches}.

Epistemic uncertainty is only present in ML-based approaches and arises from the model's lack of knowledge about the problem. It can manifest as variance in the model's predictions using the same training dataset. For example, different training runs of a deep neural network may yield different model weights, leading to varied predictions, which illustrates epistemic uncertainty. It is anticipated that adding more training data can eventually constrain the model weights, thereby reducing epistemic uncertainty~\cite{abdar2021review, gawlikowski2021survey, mena2021survey, he2023survey}. However, this process can become computationally expensive if more training data instances are added indiscriminately, and there is no assurance of improved predictive performance if the added training data instances are of low quality or do not provide meaningful information to enhance the model.

Thus, the proposed ML-based software vulnerability patch data curation process aims to address both informativeness and quality issues by measuring both epistemic and aleatoric uncertainties.This dual uncertainty measurement, along with our EHAL heuristic, defined in Section~\ref{sec:eval:activelearn}, aims to enhance the efficacy of ML-based vulnerability data curation.

\section{RQ1: What UQ techniques are capable of assessing data quality and usefulness changes in software vulnerability patch datasets?}

Researchers in engineering and scientific domains increasingly use uncertainty quantification to address measurement errors and model uncertainties~\cite{he2023survey,smith_uncertainty_2013}. While UQ studies provide numerous approaches to measure uncertainty, they also reveal that specific datasets and data types introduce domain-specific challenges and there is no one-fits-all uncertainty estimate~\cite{sinha2024domain, mucsnyi2024benchmarking}. For instance, recent studies emphasize the presence of evolving distributions, noise, and the need for domain-specific UQ methods tailored to structured representations and unique patterns in source code data~\cite{hu_codes_2023,li2021estimating}. In this first RQ, we aim to identify the most suitable UQ approximation techniques for addressing the unique noise and challenges associated with software vulnerability data curation. Applying UQ to vulnerability datasets is particularly challenging because the domain of vulnerability patches and descriptions is unique and therefore it is not possible to determine in advance which UQ method will perform best. Vulnerability patches listed in public datasets vary significantly in quality and utility, necessitating precise UQ measurements to account for these variations.

There are numerous approaches for quantifying epistemic and aleatoric uncertainty. 
Our objective is to identify suitable UQ approaches for patch data curation, focusing on neural network-based models, as they dominate ML-based curation tasks. We prioritize methods that integrate seamlessly without architectural modifications and can handle data distribution shifts common in vulnerability patches~\cite{li2021estimating}.

Probabilistic UQ methods are categorized into Bayesian and Frequentist approaches~\cite{hullermeier2021aleatoric}. While conformal prediction, a popular Frequentist method, assumes data exchangeability and offers only marginal coverage guarantees~\cite{vovk_algorithmic_2005, foygel_barber_limits_2021}, Bayesian approaches provide more flexible uncertainty estimation. However, full Bayesian networks are computationally prohibitive. Instead, we adopt Bayesian approximations, specifically Monte Carlo Dropout and Variational Inference, which deliver instance-level uncertainty efficiently and integrate well with existing models.

To answer the question of which UQ approaches are suitable for the specific domain of vulnerability patch datasets, we first introduce three UQ approximation techniques in Section III-A. Following that, Section III-B discusses two different assumptions underpinning the representation of vulnerability data, specifically whether it is homoscedastic or heteroscedastic. Subsequently, we examine metrics that can effectively gauge the accuracy of the measured uncertainty in Section III-C. Finally, we discuss the evaluation experiment setup and the results in Section III-D and Section III-E, respectively.

\subsection{UQ Approximation Techniques}
\label{sec:uq:tech}
Bayesian neural networks provide a probabilistic framework for UQ, where we have a natural representation of uncertainty, i.e., the probability distribution \(p(y|x,D,\theta)\) where \(D\) represents the training data, \(\theta\) the model parameters, \(x\) a data instance, and \(y \) the predicted label. Because Bayesian neural networks are computationally expensive~\cite{jospin2022hands}, in practice, we usually approximate Bayesian neural networks through sampling of the weights of the neural networks trained using gradient descent. Two prominent sampling approaches are Monte Carlo Dropout and Model Ensemble~\cite{gal2016dropout, rahaman2021uncertainty}. As a baseline commonly used in the literature, we also include the Vanilla UQ model that is derived directly from the training of the neural network~\cite{ovadia_can_2019}.
 
\subsubsection{Vanilla}
Deep learning models targeting software engineering applications, e.g., CodeBERT, PLBART, and CommitBART~\cite{feng2020codebert, ahmad2021unified, liu2022commitbart}, are multi-layer neural networks commonly trained with gradient descent, which can offer a prediction confidence as output. The prediction confidence is a point estimate, e.g., as $p(y|x, D) = \max\limits_{\theta} p(y|x, D, \theta)$, which is often treated as a simple and straightforward UQ solution~\cite{hin2022linevd, wang2020combining, wang2021patchdb}. 

\subsubsection{Monte Carlo Dropout}
Monte Carlo Dropout is a popular regularization technique used in training neural networks, where a neural network layer's output is connected to a dropout layer. With it, we drop, i.e, zero out, the activation of the connected neurons at probability $P_d$ during training. When dropouts are used for UQ, we also turn on dropout during inference time, which provides a sampling of a single set of neural network weights~\cite{gal2016dropout}. In essence, with Monte Carlo Dropout, we approximate probability distribution $p(y|x, D, \theta)$ with a discrete probability distribution, i.e., $p(y|x, D, \theta_i)$, $i=0, 1, 2, \ldots, T-1$ by sampling multiple sets of neural network weights with $T$ stochastic passes.

\subsubsection{Model Ensemble}
Model Ensemble is another method to sample neural network parameters. It begins with an ensemble, a set of neural networks that are, independently, initialized with their parameters (network weights) and trained. A $T$ ensemble of neural networks would provide $T$ samples of network parameters. This ensemble gives us $p(y|x, D, \theta_i)$, $i=0, 1, 2, \ldots, T-1$, which serves as an approximation to  $p(y|x, D, \theta)$. This method requires neither modification of the neural network nor the presence of dropout layers in an existing model. However, it does impose a greater training cost when compared to the Monte Carlo Dropout approach.

\subsection{\Hmsc vs. \Htsc Models}
\label{sec:setup:homohetero}
Next, we explore another dimension: how we model the distribution of noise in the changeset data, where each instance is represented by a set of features computed from the changesets. We examine two modeling choices: \hmsc and \htsc models.

\subsubsection{\Hmsc Models}
\label{sec:setup:homo}
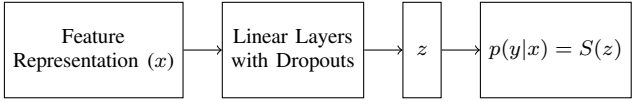
\begin{figure}[!htbp]
    \begin{adjustbox}{max width=\columnwidth}
    \begin{tikzpicture}

    \tikzset{        
        node distance=0.3in and 0.2in,
        input/.style={rectangle, draw, solid, black, align=center,
            font=\footnotesize, minimum height=0.5in, minimum width=0.2in, rounded corners=0pt},
        layer/.style={rectangle, draw, solid, black, align=center,
            font=\footnotesize, minimum height=0.5in, minimum width=0.2in, rounded corners=0pt},
        dropouts/.style={rectangle, draw, solid, black, drop shadow, fill=white,
            align=center, font=\footnotesize, minimum height=0.5in, minimum width=0.2in, rounded
            corners=3pt},
        output/.style={rectangle, draw, solid, black, align=center,
            font=\footnotesize, minimum height=0.5in, minimum width=0.2in, rounded corners=0pt},
    }

    \node[input](in){Feature\\Representation ($x$)};
    \node[layer, right=of in](mlp){Linear Layers \\ with Dropouts};
    \node[output, right=of mlp](z){$z(x)$};
    \node[output, right=of z](p){$p(y|x) = S(z)$};
    \draw[->](in) -- (mlp);
    \draw[->](mlp) -- (z);
    \draw[->](z) -- (p);
    \end{tikzpicture}
    \end{adjustbox}
    \caption{Design of the \hmsc model, where $z \in \mathbb{R}^N$, and $N=2$.}
    \label{fig:model:mlp}
\end{figure}

In \hmsc models, we assume all changeset instances are identically distributed, e.g., as a Gaussian distribution with a common mean and variance for all changesets. The classification model is therefore \hmsc as the variance is constant across all changesets. We can create a \hmsc model as follows: given the feature representations of changesets, a multi-layer perceptron (Figure~\ref{fig:model:mlp}) classifies changesets into one of two classes: vulnerability or non-vulnerability patch. The output layer has two neurons representing the outputs for these two classes. An output is commonly called a logit (denoted as \( z \)). When passing a logit through the \texttt{softmax} function, i.e., \( S(\cdot) \), we obtain a discrete probability distribution over the two classes, denoted as \( p(y|x) \) where \( x \) is the feature vector and \( y \) the random variable for the label.

\subsubsection{\Htsc Models}
\label{sec:setup:hetero}
\Hmsc models' assumption that changesets are identically distributed is unlikely to hold. For instance, changesets from different software projects or those from different development phases of a single project are likely to have different characteristics. This may result in significantly different data distribution in (the specific features of interest of) different changesets. \Htsc models although more complex may be more suitable for changeset data.

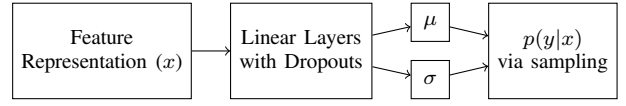
\begin{figure}[!htbp]
    \centering
    \begin{tikzpicture}

        \tikzset{        
            node distance=0.3in and 0.2in,
            input/.style={rectangle, draw, solid, black, align=center,
                font=\footnotesize, minimum height=0.5in, minimum width=0.2in, rounded corners=0pt},
            layer/.style={rectangle, draw, solid, black, align=center,
                font=\footnotesize, minimum height=0.5in, minimum width=0.2in, rounded corners=0pt},
            dropouts/.style={rectangle, draw, solid, black, drop shadow, fill=white,
                align=center, font=\footnotesize, minimum height=0.5in, minimum width=0.2in, rounded
                corners=3pt},
            soutput/.style={rectangle, draw, solid, black, align=center,
                font=\footnotesize, minimum height=0.2in, minimum width=0.2in, rounded corners=0pt},
            output/.style={rectangle, draw, solid, black, align=center,
                font=\footnotesize, minimum height=0.5in, minimum width=0.2in, rounded corners=0pt},
        }

        \node[input](in){Feature\\Representation ($x$)};
        \node[layer, right=of in](mlp){Linear Layers \\ with Dropouts};
        \node[soutput, right=of mlp, yshift=0.15in](mu){$\mu(x)$};
        \node[soutput, right=of mlp, yshift=-0.15in](sigma){$\sigma(x)$};
        \node[output, right=of mu, yshift=-0.15in](prob){$p(y|x)$\\via sampling};
        \draw[->](in) -- (mlp);
        \draw[->](mlp) -- (mu);
        \draw[->](mlp) -- (sigma);
        \draw[->](mu) -- (prob);
        \draw[->](sigma) -- (prob);
    \end{tikzpicture}
    \caption{Design of the \htsc model. It has two output heads for $\mu \in \mathbb{R}^n$ and $\sigma \in {\mathbb{R}^+}^n$, where $n=2$ is the number of classes.}
\label{fig:model:mlp2h}
\end{figure}

In this paper, we realize \htsc models via multi-output-head neural networks~\cite{kendall_what_2017}. We illustrate the model in Figure~\ref{fig:model:mlp2h}. For changeset $x$, we assume the output logit follows a Gaussian distribution, i.e., $z \sim \mathcal{N}(\mu, \sigma^2)$ (as shown in Figure~\ref{fig:model:mlp2h}). This is a model with dual output head, one representing $\mu$ and the other $\sigma$. As a \htsc model, given $N$ changesets, we obtain $N$ Gaussian distributions.

\subsection{Uncertainty Measures}
\label{sec:uq:measure}
Uncertainty is naturally captured as a probability distribution $p(y|x, D, \theta)$. However, because it can be challenging to compare probability distributions to gauge different level of uncertainty, we sometimes desire a {\em ``one-number''} quantity to indicate the uncertainty of our prediction although such a quantity do not fully capture it.

Two frequently referenced uncertainty quantities are predictive entropy and mutual information~\cite{depeweg2018decomposition, malinin2018predictive, smith2018understanding}. Via $T$ samples on model weights $w$, e.g., from $T$ stochastic passes via Monte Carlo Dropout or from $T$ models via Model Ensemble, we estimate 
$p(y|x, D) \approx \bar{p}(y|x, D) = \frac{1}{T} \sum_{\theta} p(y|x, D, \theta)$. Then we compute predictive entropy as:

\begin{align}\label{eq:uq:entropy}
    \mathcal{H}(y|x, D) = - \sum\limits_{c} & p(y|x, D) \log p(y|x, D)
\end{align}
and mutual information as:

\begin{align}\label{eq:uq:mi}
    \mathcal{I}(y|x, D)  &= \mathcal{H}(y|x, D) - \mathbb E_{p (\theta|D)} \mathcal{H}(y|x,\theta) 
\end{align}
where: 
\begin{align}\label{eq:uq:data}
    &\mathbb E_{p (\theta|D)} \mathcal{H}(y|x, \theta) \nonumber \\
        &= - \frac{1}{T}\sum\limits_{c}\sum\limits_{t=1}^{T} p(y=c | x, D, \theta_t) \log p(y=c | x, D, \theta_t)
\end{align}
where $c$ is for all possible classes, $D$ the set of training data, and $\theta$ the model parameters. 

\ronerevisiontext{
There is often a need to disentangle aleatoric and epistemic uncertainties~\cite{depeweg2018decomposition, hullermeier2021aleatoric}, as this work also demonstrates. The Law of Total Variance states that total variance can be decomposed into two components,
\begin{equation}
    Var(y) = E[Var[y|x]] + Var[E[y|x]]
\end{equation}
where, in this study, $x$ represents a {\em single} commit patch and $y$ its vulnerability status. A patch identification model, a neural network, approximates the function $y = \phi(x)$.
We treat $Var(y)$ as a representation of total uncertainty. Since $E[y|x]$ is expected to average out the noise, the variance of $E[y|x]$ can be attributed to the model's lack of knowledge during inference. As such, we consider $Var[E[y|x]]$ an estimate of epistemic uncertainty. Subtracting the epistemic component leaves the aleatoric uncertainty. Accordingly, we regard $E[Var[y|x]]$ as an estimate of aleatoric uncertainty. }
Given this understanding, in the context of \hmsc models, the literature typically refers to the entropy defined in equation~\eqref{eq:uq:entropy} as a measure of total uncertainty, the quality defined in equation~\eqref{eq:uq:data} as a measure of data or aleatoric uncertainty, and the mutual information estimated in equation~\eqref{eq:uq:mi} as a measure of model or epistemic uncertainty~\cite{depeweg2018decomposition, malinin2018predictive, smith2018understanding}.
In the context of \htsc models, we obtain two probabilistic distributions, one representing aleatoric uncertainty, $p_{ale}(y|x)$, estimated from the expectation of $\sigma$, and the other representing epistemic uncertainty, $p_{epi}(y|x)$, computed using the variance of $\mu$~(see Figure~\ref{fig:model:mlp2h}). With these two distributions, we can compute two entropy values, denoted as $\mathcal{H}_{ale}(y|x)$ and $\mathcal{H}_{epi}(y|x)$ according to equation~\eqref{eq:uq:entropy}, which correspond to aleatoric and epistemic uncertainties, respectively.
A detailed explanation of the application of the Law of Total Variance to separate total uncertainty into epistemic and aleatoric uncertainty can be found in the machine learning literature~\cite{depeweg2018decomposition, hullermeier2021aleatoric, valdenegro2022deeper}.

 \subsection{Experiment Setup}
\label{sec:uq:setup}

Table~\ref{tab:setup} provides a summary of the conditions for the experiment. In total, we utilize three UQ approximation techniques: Vanilla, Monte Carlo Dropout, and Model Ensemble (Section~\ref{sec:uq:tech}). These three techniques are examined alongside two data modeling approaches, \hmsc and \htsc (Section~\ref{sec:setup:homohetero}), resulting in six design combinations for our data curation and quality ranking algorithms. To mitigate threats to the validity of the study, we apply these techniques to two independently curated datasets using two separate feature extraction methods. In the following text, we describe the experimental setup, including the datasets, data features, and procedures.

\begin{table}[tb]
    \centering
    \caption{Summary of the experimental conditions.}\label{tab:setup}
    \small
    \begin{adjustbox}{max width=\columnwidth}
    \begin{tabular}{p{0.28\columnwidth}|p{0.62\columnwidth} }
        \toprule
        {\bf Design Factor} & {\bf Values} \\
        \midrule
        Data Modeling & \Hmsc; \Htsc\\
        UQ Approximation & Vanilla; Monte Carlo (MC) Dropout; Model Ensemble \\
        Datasets & SAP~\cite{ponta_manually-curated_2019}; VCMatch~\cite{wang_vcmatch_2022} \\
        Data Features & Embeddings (using CodeBERT~\cite{feng2020codebert}); Manually-crafted features (from PatchScout~\cite{tan_locating_2021} and VCMatch~\cite{wang_vcmatch_2022}) \\
\bottomrule
    \end{tabular}
    \end{adjustbox}
\end{table}

\subsubsection{Quality Metrics for UQ}
\label{sec:uq:metrics}
In order to evaluate the quality of the proposed UQ techniques, we rely on the Brier Score metric, which is commonly used for this purpose~\cite{lakshminarayanan2017simple,ovadia_can_2019,huseljic2021separation,rahaman2021uncertainty}, and on F1-Score, which is a popular classification metric:

\noindent
Brier Score (BS) is defined as $\text{BS}(F, Y) = \frac{1}{N}\sum_{t=1}^{N}\sum_{i=1}^{N_c} (p_{ti} - y_{ti})^2$. For classification tasks, given data instance $x_t \in X$, while $y_{ti} $ is the $i$-th element of one-hot encoded label of $y_t \in Y$, $p_{ti}$ is the predicted probability for class $i$ according to classifier $F$. Brier Score is a strictly proper scoring rule, i.e., a {\em more accurate model always produces a smaller score}.

\noindent
F1-Score is a standard classification evaluation metric. It is defined as the harmonic mean of Precision and Recall and can be computed as \(2 N_{TP}/(2 N_{TP} + N_{FP} + N_{FN})\), where \(N_{TP}\) represents the number of true positives, \(N_{FP}\) the number of false positives, and \(N_{FN}\) the number of false negatives. Because changesets are highly class imbalanced, i.e., vulnerability patches are a small minority among all patches, we choose the F1-Score for evaluation.

\subsubsection{Datasets and Feature Representation}

We use two manually-validated datasets, VCMatch and SAP. Although these datasets are relatively smaller in size, they are known for their high quality, which is crucial for the nature of our experiment. High-quality data is essential to ensure that the evaluation is not contaminated by the noise commonly present in many vulnerability patch datasets, as reported in numerous prior studies~\cite{zhou_automated_2017, cabrera_commit2vec_2021, wang_patchrnn_2021, zhou_finding_2021, zhou_spi_2021, wu_enhancing_2022, tan_locating_2021, wang_vcmatch_2022}. The VCMatch dataset contains $1,669$ researcher-validated vulnerability patches and matching CVEs from 10 OSS projects (FFmpeg, ImageMagick, Jenkins, Linux, Moodle, OpenSSL, phpMyAdmin, PHP-src, QEMU, and Wireshark)~\cite{wang_vcmatch_2022}. The SAP dataset is validated by their production software teams and contains 1,282 patches to publicly disclosed vulnerabilities affecting 205 distinct open-source Java projects referenced by SAP's software~\cite{ponta2019manually}. 
In the datasets, we need both vulnerability and non-vulnerability patches. For each vulnerability patch, we randomly sample $5,000$ non-vulnerability patches from the software projects, similar to the process performed by VCMatch and PatchScout~\cite{tan_locating_2021, wang_vcmatch_2022}.

We consider two separate feature representations of changesets: 1) manually designed features and 2) embeddings. For the former, we adopt the features proposed in PatchScout~\cite{tan_locating_2021} and also re-used by VCMatch~\cite{wang_vcmatch_2022}. These features are computed using NVD records, CVE and CWE attributes, and changesets commit messages and diffs. PatchScout categorizes these 22 features into 4 broad categories of vulnerability identifier features, vulnerability location features, vulnerability type features, vulnerability description text features. For the embeddings, we rely on a popular state-of-the-art LLM -- CodeBERT~\cite{feng2020codebert}. In this study, as in many others, we treat the last hidden states of CodeBERT's neural network as embeddings. 

\subsubsection{Evaluation Procedures}
\label{sec:setup:procedures}
We implement six deep learning models by combining three UQ techniques (Section~\ref{sec:uq:tech}) with two model architectures (Section~\ref{sec:setup:homohetero}). First, we develop Vanilla models for both \hmsc and \htsc noise assumptions (Figures~\ref{fig:model:mlp} and~\ref{fig:model:mlp2h}). Each model uses CodeBERT for feature extraction, followed by three fully connected layers with dropout. The \hmsc model applies \texttt{softmax}, while the \htsc model outputs $\mu$ via \texttt{ReLU} and $\sigma$ via \texttt{softplus} to ensure positivity. We extend these base models to support Monte Carlo Dropout by enabling stochastic dropout during training and inference, and Model Ensemble by training five independent models and averaging their predictions. 

We use the Adam optimizer during training. For \hmsc models, we train with the negative log likelihood loss function (i.e., mathematically, the cross-entropy loss in two-class classification), and for \htsc models, we train with a stochastic negative log likelihood loss function introduced by Kendall and Gal~\cite{kendall_what_2017}.

We tune hyperparameters using a validation dataset and monitor training with early stopping, halting after five epochs of no improvement in validation loss. Each layer contains 300 neurons with a dropout rate of 0.1. The model checkpoint with the lowest validation loss is selected for evaluation.

To answer RQ1, we design and carry out a series of numerical experiments. For each experiment, we follow standard training, validation and testing protocols. Unless otherwise stated, the ratio of training and test datasets is $0.8:0.2$. From the training dataset, 10\% is set aside as a validation dataset. During training, the majority class, i.e., non-vulnerability patches, are under-sampled to have a balanced training dataset, while during evaluation the test dataset is kept as is.

 \subsection{Evaluation Results}
\label{sec:uq:results}

We present the results organized along different themes focused on providing a holistic answer to RQ1.

\subsubsection{Comparison of UQ Techniques}
\label{sec:eval:uqquality}

\begin{figure*}[!htbp]
    \centering
\subfloat[]{\includegraphics[width=0.32\textwidth]{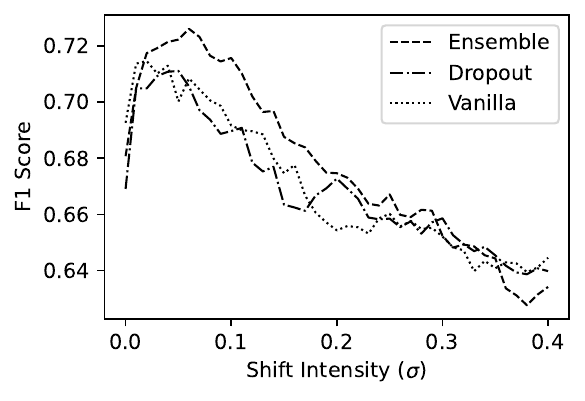}\label{fig:uq:si:im1:htsc:psvcm:f1}
    }
    \hfil
    \subfloat[]{\includegraphics[width=0.32\textwidth]{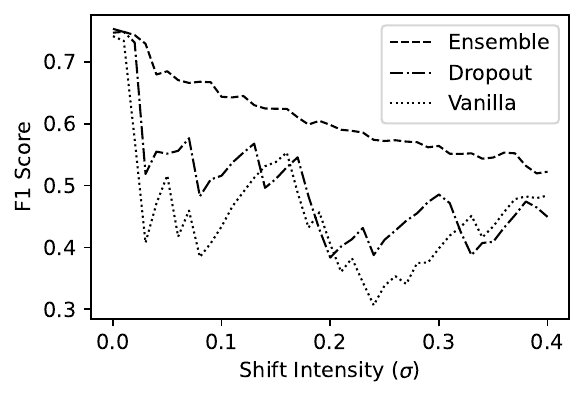}\label{fig:uq:si:im1:htsc:bertvcm:f1}
    }
    \hfil
    \subfloat[]{\includegraphics[width=0.32\textwidth]{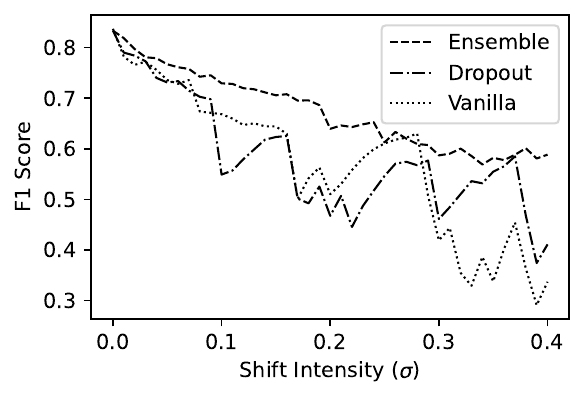}\label{fig:uq:si:im1:htsc:bertsap:f1}
    }

    \subfloat[]{\includegraphics[width=0.32\textwidth]{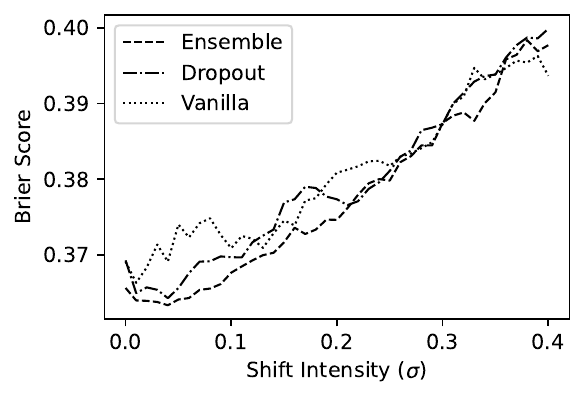}\label{fig:uq:si:im1:htsc:psvcm:brier}
    }
    \hfil
    \subfloat[]{\includegraphics[width=0.32\textwidth]{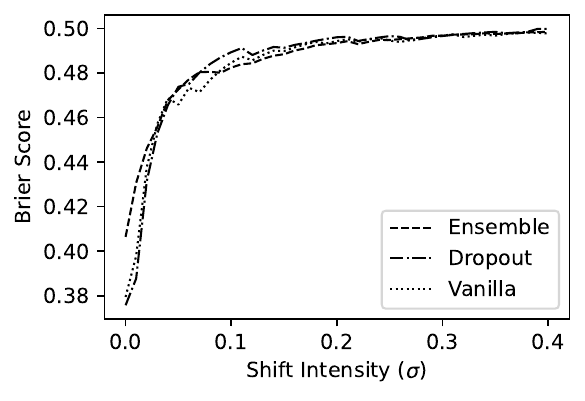}\label{fig:uq:si:im1:htsc:bertvcm:brier}
    }
    \hfil
    \subfloat[]{\includegraphics[width=0.32\textwidth]{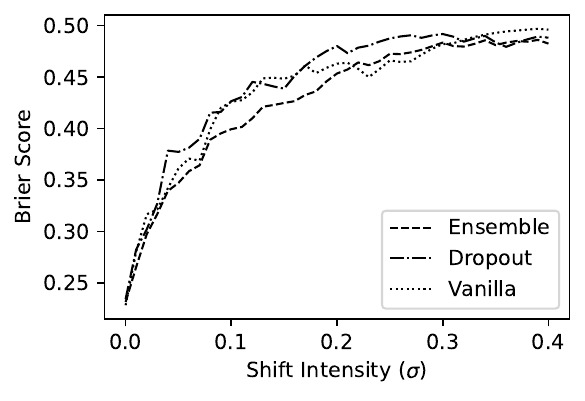}\label{fig:uq:si:im1:htsc:bertsap:brier}
    }
    
    \caption{F1-score and Brier Score vs. dataset quality changes of \htsc models. Subfigures~\protect\subref{fig:uq:si:im1:htsc:psvcm:f1} to~\protect\subref{fig:uq:si:im1:htsc:bertsap:f1} show the F1-score obtained from
    three experimental settings: (a) VCMatch dataset with PatchScout features; (b) VCMatch dataset with CodeBERT features;
    and (c) the SAP dataset with CodeBERT features; Subfigures~\protect\subref{fig:uq:si:im1:htsc:psvcm:brier} to~\protect\subref{fig:uq:si:im1:htsc:bertsap:brier} show the corresponding Brier Score for the three experiments.}
    \label{fig:uq:si:im1}
\end{figure*}

Among numerous UQ approximation methods in the literature, in this paper, we consider three of the most popular: Vanilla, Model Ensemble, and Monte Carlo Dropouts. In order to examine which of these three methods has the best potential to support patch data curation, we train and test these methods with increased dataset quality shift. The dataset quality shift is an artificially induced quality degradation in the data. To this end, we add Gaussian noise to each of the features by drawing a random sample $n$ from a Gaussian distribution with zero mean and diagonal covariance, i.e., from $\mathcal{N}(0, \Sigma)$ where $\Sigma$ is the covariance matrix. We let $\Sigma = \sigma I $ where $I$ is the identity matrix, i.e., the added noise is independent for different features in the feature vector $x$ of each changeset. We refer to $\sigma$ as the shift intensity that we vary.
Figure~\ref{fig:uq:si:im1} shows the F1-score and Brier Score for increasing dataset shift intensity. The overall trend is that when the quality shift intensity increases, the model's predictive performance and uncertainty get worse (lower F1-score and higher Brier Score). These observations demonstrate that the UQ models respond appropriately to the degraded quality in the data, i.e., when data quality degrades the models suffer from decreased predictive performance (i.e., F1-score) along with increased prediction uncertainties (i.e., Brier Score). The results in Figure~\ref{fig:uq:si:im1} are obtained via \htsc models, while similar experimental results for \hmsc models show the same trends. 

While all three UQ approximations capture the overall trend of the data quality shift, we observe that Model Ensemble generally yields the highest predictive performance ($1.5\% - 12.4\%$ gain on F1 score) and the most accurate UQ estimation ($\sim 1\%$ reduction in Brier Score). These observations are consistent with those obtained from similar experiments in the literature~\cite{lakshminarayanan2017simple, ovadia_can_2019, rahaman2021uncertainty}, which also show Model Ensemble to be a most robust UQ approximation method that can be leveraged for data curation of software vulnerability patches. Therefore, we conclude that this UQ model is likely to be most beneficial to curate vulnerability patch datasets.

\subsubsection{Separating Epistemic from Aleatoric Uncertainty}
\label{sec:eval:knowledge}

When selecting training data for a machine learning model, two key factors are the quality and quantity of data. While high-quality data is preferred, low-quality data can still be useful when no other alternatives are available, as it may help the model gain some understanding of the problem space. Therefore, the decision to include a patch in the dataset should be based on epistemic uncertainty rather than just data quality, i.e., aleatoric uncertainty. A Uncertainty Quantification (UQ) technique that distinguishes between epistemic and aleatoric uncertainties can indicate whether a patch can contribute to the model's understanding. This is because epistemic uncertainty reflects the model's knowledge level; high epistemic uncertainty suggests a lack of knowledge about the problem space.

In this section, we aim to ascertain whether the best UQ model we evaluated above, i.e., the \htsc Model Ensemble, can serve as a reliable base for accurately outputting the disentangled uncertainties, i.e., the epistemic and aleatoric uncertainties.We demonstrate that epistemic and aleatoric uncertainties behave as expected when curating patch data. In Table~\ref{tab:epivsale}, we set aside 80\% of patches as training data, train the patch identification model with increasing training dataset sizes (60\%, 80\%, and 100\% of the training data), and then estimate the epistemic and aleatoric uncertainties on the test dataset (the remaining 20\% of the patches). The table shows that as we increase the training data size, there is a significant reduction in epistemic uncertainty, i.e., a 1.5\% to 9.3\% decrease in epistemic uncertainty when the training data increased from 60\% to 80\% and then to 100\%. In contrast, the aleatoric uncertainties saw smaller changes. Since an increase in the training data size indicates an improvement in the model's knowledge about the problem space, which is captured by a reduction in epistemic uncertainty, it follows that for data curation of software vulnerability patches, a UQ technique that disentangles epistemic and aleatoric uncertainties is required.

\begin{table}[tb]
    \caption{Epistemic Uncertainty vs. Aleatoric Uncertainty as Training Data Increases}
    \label{tab:epivsale}
    \begin{adjustbox}{max width=\columnwidth}
    \begin{tabular}{p{0.35in} r r r r}
        \toprule
        Data
            & $\overline{\mathcal{H}}_{epi}$
            & $\Delta_{epi}\%$
            & $\overline{\mathcal{H}}_{ale}$
            & $\Delta_{ale}\%$ \\
        \midrule
        60\% & 0.794 &                                    0.0\% & 0.817 &                                    0.0\% \\
        80\% & 0.782 & $\frac{0.794-0.782}{0.794}\approx 1.5\%$ & 0.819 & $\frac{0.817-0.819}{0.817}\approx-0.2\%$ \\
       100\% & 0.709 & $\frac{0.782-0.709}{0.782}\approx 9.3\%$ & 0.790 & $\frac{0.819-0.790}{0.819}\approx 3.5\%$ \\
       \bottomrule
    \end{tabular}
    \end{adjustbox}
\end{table}

\subsubsection{\Htsc vs. \Hmsc Models}

Finally, we pose the question of whether modeling the noise in software patch data as \htsc offer any advantage over modeling it as \hmsc. To optimize computational resources, we focused on the \htsc Model Ensemble, the best-performing UQ approximation, for comparison against \hmsc UQ techniques.The results in Table~\ref{tab:uq:uq_comp} are obtained by averaging over all data quality shift intensities (from 0 to 0.4 with a step size of 0.1). The results show that the \htsc model offers a significant advantage in predictive performance (higher F1-score) and superior or similar UQ quality (lower or nearly equal Brier Score). Given this observation, we conclude that for vulnerability patches, the UQ measures offered by the \htsc models are more accurate than the \hmsc models, due likely to the fact that the data distribution in vulnerability patches can vary between projects (and within a project over time), which is more accurately captured by \htsc models.

\begin{table}[tb]
    \centering
    \caption{Comparison of UQ modeling approaches}\label{tab:uq:uq_comp}
    \begin{adjustbox}{max width=\columnwidth}
    \begin{tabular}{p{1.75in}|c|c c}
        \toprule
            {\bf UQ Model \& Assumption} & {\bf F1-score}          & {\bf Brier Score}     \\
        \midrule
        \Htsc Ensemble & $\mathbf{0.687 \pm 0.017}$ & $\mathbf{0.362 \pm 0.002}$ \\
        \Hmsc Ensemble & $0.652 \pm 0.031$ & $0.395 \pm 0.012$ \\
        \Hmsc Dropout & $0.651 \pm 0.029$ & $0.396 \pm 0.012$ \\
        \Hmsc Vanilla & $0.650 \pm 0.029$ & $0.396 \pm 0.012$ \\
        \bottomrule
    \end{tabular}
    \end{adjustbox}
\end{table}

\begin{mdframed}
\textit{
        \begin{enumerate}[leftmargin=0.2in]
\item From the three UQ models we considered, Model Ensemble produces the most accurate UQ estimation in response to artificial data quality degradation.  
\item To effectively assess the knowledge gained by a machine learning-based automated patch curation, it is crucial to separate epistemic and aleatoric uncertainty. Epistemic and aleatoric uncertainty are distinguishable during automated patch data curation.
            \item \Htsc models provide more accurate UQ estimation for patch data curation.
        \end{enumerate}
    }
\end{mdframed}

\section{RQ2: Can we use UQ to improve automatically curated vulnerability patches by selecting high quality and highly usable security patches?}
\label{sec:eval:activelearn}
Based on RQ1, we design and evaluate the data curation system. First, we assess the algorithm’s ability to predict patch vulnerability. Second, we examine its impact on vulnerability prediction, a key downstream application (Figure~\ref{fig:eval:setting:rq2}).

We begin with a small set of high-quality seed patches $(X_s, Y_s)$ to train a UQ model. Next, we curate patches from a pool $X_p$ of potentially low-quality or project-specific patches. To prevent overlap, we ensure $X_p \cap X_s = \emptyset$. The curation process involves two steps: running inference on $X_p$ and applying the EHAL heuristic to select informative, high-quality patches.

\begin{figure}[!htbp]
    \centering
    \begin{adjustbox}{max width=\columnwidth}
    \begin{tikzpicture}

        \tikzset{        
        node distance=0.3in and 0.5in,
        every node/.style={font=\footnotesize},
        input/.style={rectangle, draw, solid, black, align=center,
            font=\footnotesize, minimum height=0.5in, minimum width=0.2in, rounded corners=0pt},
        procedure/.style={rectangle, draw, solid, black, align=center,
            font=\footnotesize, minimum height=0.5in, minimum width=0.2in, rounded corners=0pt},
        output/.style={rectangle, draw, solid, black, align=center,
            font=\footnotesize, minimum height=1.1in, minimum width=0.2in, rounded corners=0pt},
        }

        \node[input, anchor=west](pool) at (0, 0) {Pool of Patches \\($X_p$ where $X_p \cap X_s = \emptyset$)};
        \node[input, above=of pool, anchor=west](seed) at (0, 0.3in){Seed \\Patches\\ ($X_s, Y_s$)};
        \node[procedure, above=of pool, anchor=east](model) at ($(pool.east) + (0, 0.3in)$){UQ \\Model};

        \node[procedure, minimum height=1.1in, right=of pool, yshift=0.3in](uq) {UQ\\measures:\\$U_p=$\\$U_{ale}$,\\$U_{epi}$,\\$P(\hat{Y}_p|X_p)$};

        \node[output, right=of uq](patches) {Selected\\Patches:\\$X_c \subseteq X_p$};
        
 		\draw[->](seed) -- (model) node[midway, above] {Train};
        \draw [->](model) -- ($(model.east)!0.5!(uq.west)$) -- (uq) node[midway, below, xshift=-0.1in] {Inference};
        \draw [->](pool) -- ($(pool.east)!0.5!(uq.west)$) -- (uq);
        \draw[->](uq) -- (patches) node[midway, above, text width=0.3in, align=center] {Select \\via EHAL};
    \end{tikzpicture}
    \end{adjustbox}

    \caption{Evaluation setting for RQ2.}
    \label{fig:eval:setting:rq2}
\end{figure}

\subsection{Automatic Curation of Vulnerability Patches with UQ}

Given an accurate estimation of epistemic and aleatoric uncertainty for patch data curation, as described in RQ1, it is still an open problem of how we use these two quantities to improve patch data curation. 
In this RQ2, we focus on the practical aspects of using UQ in ML-based patch data curation in order to obtain higher quality datasets. To this end, we design the EHAL Heuristic (Algorithm~\ref{alg:vuldataehal}) that actively and selectively adds patches based primarily on high epistemic uncertainty to include in the dataset.
\ronerevisiontext{The algorithm relies on a pre-trained UQ model, which takes as input a pool of commit patches and outputs a selected subset. These input patches may be either labeled or unlabeled. For unlabeled patches, the algorithm identifies candidates for annotation, while for labeled ones, it selects instances that are both informative and reliable for training.
}

Recall that a high epistemic uncertainty of a patch indicates that the model lacks knowledge about this patch, thus the patch is valuable and should be added to the dataset. However, in RQ1 we also observed that a significant degradation of data quality can lead to reduced model performance. Since aleatoric uncertainty is an indicator of the noise in the data, we should avoid those patches that have high aleatoric uncertainty. Therefore, our EHAL algorithm rests on a simple assumption: we should select patches with high epistemic uncertainty, while we should avoid (or not select) those with high aleatoric uncertainty.

Instead of leveraging a formula that combines both epistemic and aleatoric uncertainties, the EHAL heuristic takes a ``select-and-then-reject'' approach. \ronerevisiontext{We opted for this strategy over scalar combinations (e.g., weighted sums of uncertainties) to avoid arbitrary tradeoff calibration. Empirically, we found this approach simpler to tune and less sensitive to outliers in either uncertainty dimension.} As shown in Algorithm~\ref{alg:vuldataehal}, it first identifies patch instances with the highest epistemic uncertainty as candidates, but rejects them if they are among the patches with the highest aleatoric uncertainty, i.e., it rejects those on the lowest extreme of the quality spectrum. The name of the heuristic is drawn from this idea as well: Epistemic High Aleatoric Low (EHAL).

More specifically, the algorithm’s main function, \texttt{DataCurationWithEHAL} (Lines 1–8), curates $n$ patches from a candidate pool $\set{X}_{\text{pool}}$, guided by their pre-computed epistemic and aleatoric uncertainties, $\set{U}_{\text{pool}}$. It iteratively selects patches by invoking the \texttt{EHAL} heuristic (Lines 9–19), which forms the core of the algorithm. The heuristic identifies the candidate patch with the highest epistemic uncertainty using the \texttt{TopOneByEpistemicUQ} operation (Lines 12 and 17). Simultaneously, the top $n_{\text{ale}}$ patches with the highest aleatoric uncertainty are determined using \texttt{TopNByAleatoricUQ} (Lines 13 and 18), where $n_{\text{ale}}$ is a hyperparameter determined empirically. If the candidate with the highest epistemic uncertainty is also among the top aleatoric candidates, it is excluded from consideration, and the selection process iterates. This ensures that the final curated dataset includes only patches that are both highly informative and reliable, effectively balancing the trade-off between informativeness (epistemic uncertainty) and quality (low aleatoric uncertainty).

To evaluate the EHAL Heuristic, we consider two baselines. We construct a data selection heuristic completely opposite to the EHAL Heuristic, i.e., Epistemic Low Aleatoric High (ELAH). ELAH selects the patches with the lowest epistemic uncertainty, and rejects them if they are also among those with the lowest aleatoric uncertainty. In addition, we also compare our algorithm with the random baseline that selects patches randomly. Random baselines are frequently used in the active learning literature~\cite{parvaneh2022active}. By comparing with a random baseline, we can determine whether the data curation algorithm has any knowledge about data selection at all.

In our evaluation, we gradually expand the training dataset, using one of the three algorithms: the proposed one and the two baseline algorithms. For each, we divide the training data partition randomly into two parts: the initial training dataset (20\% of available patches) and the candidate training instances pool (60\% of available patches). We construct the training dataset, starting with the initial training dataset, and train a patch identification model. With the trained model, we estimate uncertainty measures, such as aleatoric and epistemic uncertainties. Based on the uncertainties, we select 10\% of instances from the candidate pool and add these to the training dataset. We then retrain the model and repeat the process. We stop when we exhaust the candidate dataset. To evaluate the model's predictive performance, we compute the F1 score on the test dataset (20\% of available patches) as the size of the training dataset increases. To ensure the reliability of the results, we repeat the process 10 times, each beginning with a random shuffling of the entire dataset.

\begin{algorithm}[tb]
	\caption{Patch Curation with EHAL Heuristic}
	\label{alg:vuldataehal}
    \small

	\DontPrintSemicolon
	\LinesNumbered

	\SetKwProg{Function}{Function}{}{end}
	\SetKwFunction{DataCurationWithEHAL}{DataCurationWithEHAL}
	\SetKwFunction{EHAL}{EHAL}
    \SetKwFunction{EHALn}{EHALn}

 \KwIn{
$\set{X}_{\text{pool}}$: Pool of candidate patches\\
        $\set{U}_{\text{pool}}$: Uncertainty values for candidate patches\\
        $n$: Number of patches to curate\\
        $n_{\text{ale}}$: Hyperparameter for the number of aleatoric instances to consider in each iteration
    }
    \KwOut{$\set{D}_{\text{new}}$: Final list of curated patches}

    \Function{\DataCurationWithEHAL{$\set{X}_{\text{pool}}$, $\set{U}_{\text{pool}}$, $n$, $n_{\text{ale}}$}}{
        $\set{D}_{\text{new}} \gets [~]$ \\
        \While{$\set{U}_{\text{pool}} \neq \emptyset$ \textbf{and} $\set{D}_{\text{new}}.\text{size}() < n$}{
            $d_{\text{epi}} \gets \EHAL(\set{X}_{\text{pool}}, \set{U}_{\text{pool}}, n_{\text{ale}})$ \\
            $\set{X}_{\text{pool}} \gets \set{X}_{\text{pool}} \setminus \{d_{\text{epi}}\}$ \\
            $\set{U}_{\text{pool}} \gets \set{U}_{\text{pool}} \setminus \{d_{\text{epi}}\}$ \\
            $\set{D}_{\text{new}}.\text{append}(d_{\text{epi}})$ \\
        }
        \Return $\set{D}_{\text{new}}$
    }

	\vskip 1em

	\Function{\EHAL{$\set{X}_{\text{pool}}$, $\set{U}_{\text{pool}}$, $n_{\text{ale}}$ }}{
        $\set{X}_{\text{candidate}} \gets \set{X}_{\text{pool}}$ \\
        $\set{U}_{\text{candidate}} \gets \set{U}_{\text{pool}}$ \\
$d_{\text{epi}} \gets \text{TopOneByEpistemicUQ}(\set{X}_{\text{candidate}}, \set{U}_{\text{candidate}})$\\
        $\set{D}_{\text{ale}} \gets \text{TopNByAleatoricUQ}(\set{X}_{\text{candidate}}, \set{U}     _{\text{candidate}}, n_{\text{ale}})$ \\
        \While{$d_{\text{epi}} \in \set{D}_{\text{ale}}$}{
            $\set{X}_{\text{candidate}} \gets \set{X}_{\text{candidate}} \setminus \{d_{\text{epi}}\}$ \\
            $\set{U}_{\text{candidate}} \gets \set{U}_{\text{candidate}} \setminus \{d_{\text{epi}}\}$ \\
$d_{\text{epi}} \gets \text{TopOneByEpistemicUQ}(\set{X}_{\text{candidate}}, \set{U}_{\text{candidate}})$\\
            $\set{D}_{\text{ale}} \gets \text{TopNByAleatoricUQ}(\set{X}_{\text{candidate}}, \set{U}_{\text{candidate}}, n_{\text{ale}})$ \\
        }
        \Return $d_{\text{epi}}$
	}
\end{algorithm}

\begin{figure}[!t]
    \centering
    \subfloat[]{\includegraphics[width=0.9\columnwidth]{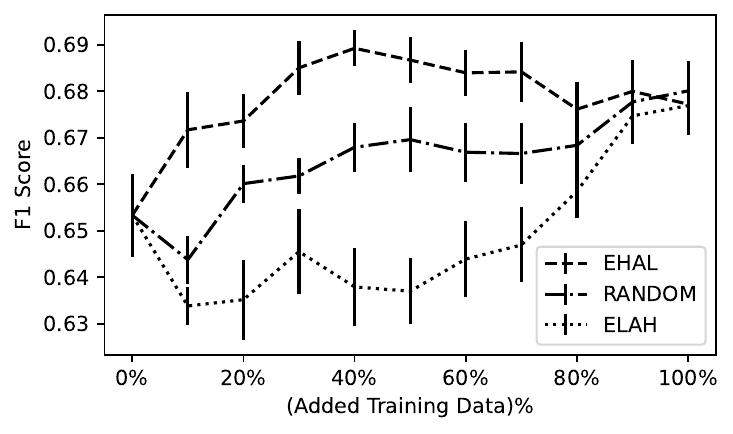}}
    \label{fig:eval:quality:ps}

    \subfloat[]{\includegraphics[width=0.9\columnwidth]{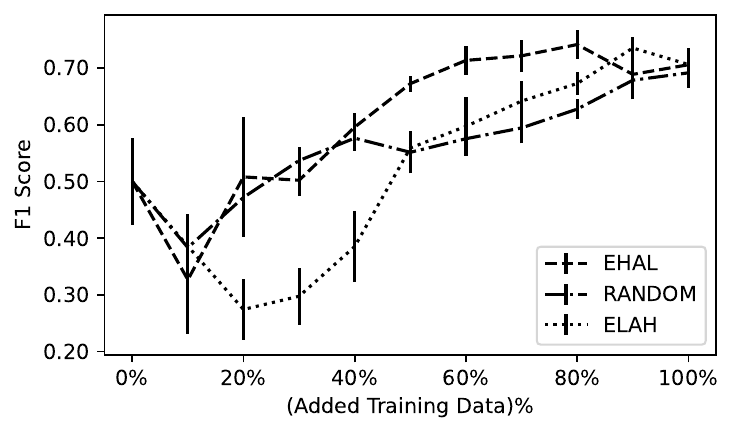}}
    \label{fig:eval:quality:sap}
    \caption{Comparison of three algorithms/heuristics for selecting training data: Epistemic High Aleatoric Low (EHAL), Epistemic Low Aleatoric High (ELAH), and random choice of data. We compute the F1-score on the: a) VCMatch dataset with PatchScout features and b) SAP dataset with CodeBERT embeddings. The error bars represent the variance from 10 independent runs while the line represents the average. }
    \label{fig:uq:active:f1}
    \vskip -1em
\end{figure}

Figure~\ref{fig:uq:active:f1} presents F1-scores obtained by \htsc models using both the VCMatch dataset with PatchScout features and the SAP dataset with CodeBERT feature representations. There are two primary observations from these results:

\paragraph{Improved Patch Curation} The results for both datasets show that the model using EHAL performs the best (F1 score) as patches are selected and added to the training dataset. The model with the ELAH baseline performs the worst, while random selection is in the middle.
This serves as evidence that the proposed algorithm can effectively select patches and improve predictive performance.
Note, however, that as the percentage of added training data approaches 100\%, there is less leeway for selection, i.e., the dataset runs out of "good" candidates, and adding more patches to the training dataset has no discernible effect. The modest gains observed in Figure~\ref{fig:uq:active:f1} are partially due to the limitations of the SAP and VCMatch datasets, which contain only 1,200 to 1,700 instances. With datasets of this size, it is challenging to observe a large effect.

\paragraph{Better Information Efficiency} As shown in Figure~\ref{fig:uq:active:f1}, the model trained using the VCMatch dataset reaches the best predictive performance with the EHAL Heuristic (Algorithm~\ref{alg:vuldataehal}) when 40\% of the patches are added to the training dataset. Since the training time is proportional to the size of the training data, the implication is that by selecting the right patches for the training dataset, we can significantly reduce the training time, i.e., we only need to train with 40\% of the data to achieve the best model performance. We observe the similar results regarding the model trained using the SAP data, although the model reaches the best performance when 80\% of the data is used for training (Figure~\ref{fig:uq:active:f1}). This also indicates the SAP datasets is of higher quality as it contains more high quality instances.  

\ronerevisiontext{
The training data ratio is an important hyperparameter in this approach, as it influences the quality of the data curation model. 
Its optimal value varies depending on the dataset and the model. Our experiments show that, for the VCMatch dataset, performance gains plateau after using approximately 40\% of the curated data, whereas for the higher-quality SAP dataset, 80\% was needed to reach maximum utility. This variability underscores the importance of tailoring the method to specific use cases, since the balance between data quality, model performance, and training-time efficiency differs across real-world applications. The data curation process is inherently iterative. A practical approach is to start with a small random sample of labeled patches, estimate a suitable ratio, and then adjust it based on model performance as more data are added.
}

\begin{mdframed}
    \noindent
    \textit{The evaluation results indicate that the data curation heuristic EHAL can effectively select patches that are high-quality and have high utility values. Using the selected patches results in improvements in the model for automatic vulnerability patch data curation with less training data.
    }
    \end{mdframed}

 \subsection{Evaluating Patch Curation's Impact on Automated Software Vulnerability Prediction}

In this section, we investigate whether our automatic patch curation algorithm, which uses the EHAL heuristic, can improve software vulnerability prediction~\cite{kalouptsoglou2023software}. This demonstrates the algorithm’s ability to select high-quality patches, thereby enhancing downstream applications and specifically improving existing vulnerability prediction models.

Automated software vulnerability prediction is a popular application area that relies on vulnerability patch data. For our experiment, we selected the LineVul vulnerability prediction algorithm~\cite{fu2022linevul}. This choice is based on three key considerations. First, LineVul is a state-of-the-art model for predicting both vulnerable functions and vulnerability-inducing lines of code~\cite{fu2022linevul}. Second, it was recently evaluated as one of the best-performing models among various vulnerability prediction approaches~\cite{steenhoek2023empirical}. Third, it utilizes the Big-Vul dataset~\cite{fan_cc_2020}, a  vulnerability dataset where each record contains either a vulnerable or non-vulnerable function. This dataset is widely used because it provides the original patches from which vulnerable functions are derived, allowing for the inference of lines deleted and added before and after security fixes, thus facilitating line-level vulnerability prediction. 
Additionally, the file-level patches in the Big-Vul dataset allow for assembling commit-level software patches, which our algorithm is designed to validate using UQ.

For this experiment, we apply the EHAL heuristic to select a subset of data from the training dataset, which we use to train the LineVul model. We then evaluate LineVul to assess two dimensions of model performance: predictive performance and computational time. For predictive performance, we examine the F1 score, as done in the LineVul study, while for computational time, we measure the training duration. Our aim is to demonstrate that our UQ-based approach can select high-quality data that improves LineVul's predictive performance and  reduces its training time.

More specifically, we carry out the experiment according to the following set of steps:
\begin{enumerate}
\item {\em Train the patch curation model.} We train a patch curation model using the VCMatch dataset and select the model that demonstrates the highest patch identification performance within the VCMatch dataset, leveraging 40\% of the training instances (see Figure~\ref{fig:uq:active:f1}). This is because both the VCMatch dataset and the Big-Vul dataset consist of C/C++ code, while the SAP dataset contains solely Java code.\item {\em Remove overlapping projects from the Big-Vul dataset.}
The Big-Vul dataset contains functions from 310 software projects, among which 8 are also in the VCMatch dataset. To eliminate the risk of data contamination, we remove these 8 software projects, namely, FFmpeg, QEMU, OpenSSL, WireShark, PHP-SRC, Moodle, ImageMagick, and Linux, from the Big-Vul dataset completely. As a result, the number of functions in the Big-Vul dataset is reduced from 188,636 to 130,449. For convenience, in the remaining discussion of this section, we refer to this filtered version as the Big-Vul dataset.\item {\em Extract commit patches from the filtered Big-Vul dataset.} We identify commit hashes associated with vulnerable functions in Big-Vul and use these hashes to extract the relevant patches.
\item {\em Partition data for LineVul training, validation, and testing.} To ensure the reliability of the results, we apply the data split method used in 10-fold cross validation to divide the commit patches in the Big-Vul dataset into the train, validation, and testing partitions, and the ratios of the commits in these three partitions are $0.8:0.1:0.1$. This procedure yields 10 distinct splits with 10 unique test partitions, ensuring no overlap of commits among the training, validation, and test partitions.
\item {\em Use the patch curation model to select LineVul training data.} We select patches from the training partition using the EHAL heuristic. To examine the effects of adding the selected security patches, we gradually increase the selected security patches, i.e., by selecting 10\%, 20\%, 30\%, $\ldots$, and 100\% of the security patches.
\item {\em Train and evaluate LineVul.} In the Big-Vul data, a vulnerable function has a number of correspondent non-vulnerable functions, all of these are indexed by the commit id of the security patch from which the vulnerable function is extracted.  This organization of the Big-Vul data allows us to assemble the training data from the selected security patches. We train a LineVul model from scratch using the functions indexed by the commit id of the selected patches, and evaluate the performance of the trained LineVul model on the test partition. The vulnerable functions in the test partition is about 6.1\%, and following the LineVul study~\cite{fu2022linevul}, we report F1 score as the performance metric. 
\end{enumerate}

\begin{figure}[!htbp]
    \centering
    \includegraphics[width=\columnwidth]{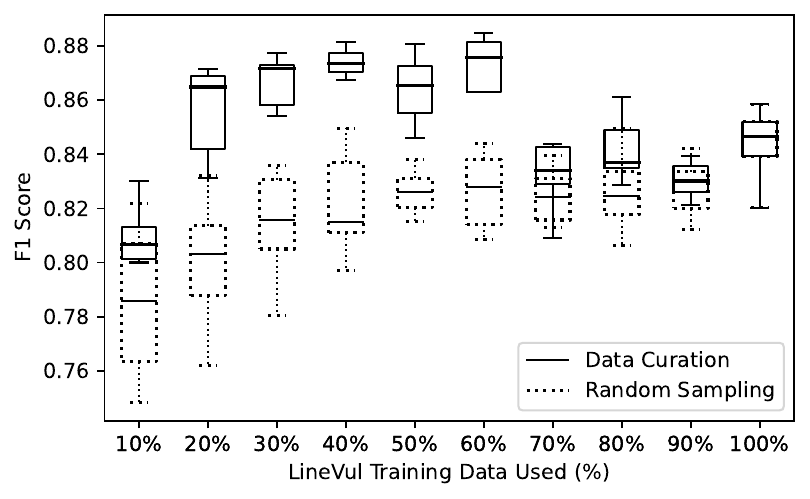}
    \caption{Performance of LineVul trained using the proposed patch curation method and the random selection method.}
    \label{fig:linevul:comparison}
\end{figure}

First, we examine the relationship between LineVul's predictive performance and the amount of training data used. As a baseline, we run the identical experiments using a random data selection method. Figure~\ref{fig:linevul:comparison} summarizes the experimental results where we show F1 score versus the amount of training data used. LineVul's predictive performance varies given different test data. The figure exhibits the variance using a standard boxplots showing min, first quartile, median, third quartile, and max. The figure shows that (a) when we use the proposed patch selection method, LineVul achieves a higher F1 score with 20-60\% of the data than with 100\% of the training data, and (b) the proposed method consistently outperforms random selection except the case when nearly 90\% or more patches are selected.

Additionally, in Table~\ref{tab:linevul:mean} we compare the mean F1 scores obtained when using 100\% of training data, using the training data selected  by the proposed data curation method, and using random selection. When comparing the proposed data curation method with no data curation, i.e, using 100\% of training data, the improvement of mean F1 score ranges from 1.66\% to 3.52\%.We observe a higher improvement of F1 score when comparing the proposed data selection with the random data selection.

\begin{table}[!htbp]
\centering
\caption{Comparing Mean F1 Scores: using 100\% Training Data (Mean F1 Score=0.843), using Proposed Data Curation Method, and Random Selection}
\label{tab:linevul:mean}
\begin{tabular}{r|r r|r r}
\toprule 
           & \multicolumn{2}{c|}{\bf \revisiontext{Mean} F1 Score} & \multicolumn{2}{c}{\bf Gain} \\
Selected~\% & Random & Curation & vs. Random  & vs. 100\% \\
\midrule
\revisiontext{20\%} & \revisiontext{0.800} & \revisiontext{0.857} & \revisiontext{7.07\%} & \revisiontext{1.66\%} \\
\revisiontext{30\%} & \revisiontext{0.818} & \revisiontext{0.868} & \revisiontext{6.02\%} & \revisiontext{2.94\%} \\
\revisiontext{40\%} & \revisiontext{0.823} & \revisiontext{0.872} & \revisiontext{6.05\%} & \revisiontext{3.52\%} \\
\revisiontext{50\%} & \revisiontext{0.825} & \revisiontext{0.858} & \revisiontext{3.98\%} & \revisiontext{1.81\%} \\
\revisiontext{60\%} & \revisiontext{0.825} & \revisiontext{0.860} & \revisiontext{4.22\%} & \revisiontext{2.08\%} \\
\bottomrule
\end{tabular}
\end{table}

Although the improvement is not large, we stress that the improvement is statistically significant. We conduct the Mann-Whitney U test to test the hypothesis that the proposed data curation method yield greater F1 scores despite the variance and the small effective size. The Mann-Whitney U test is a non-parametric test that is suitable for F1 scores, one-sided thus non-Gaussian. Table~\ref{tab:linevul:hptest} shows that p-value is far less than 5\% and we reject the null hypothesis that we obtain superior F1 scores using the proposed data curation method is due to chance. Given that we control all factors in our experiments, we conclude that the proposed method is superior, as it robustly yields higher F1 scores.

\begin{table}[!htbp]
\centering
\caption{Hypothesis Testing via Mann-Whitney U Test: Proposed Data Curation Produces Greater F1 Score
than using 100\% data and than Random Selection}
\label{tab:linevul:hptest}
\begin{tabular}{r|r|r}
\toprule 
           & \multicolumn{2}{c}{\bf p-value} \\
Selected~\% & \revisiontext{Curation vs. Random} & \revisiontext{Curation vs. 100\%} \\
\midrule
\revisiontext{20\%} & \revisiontext{$8.23e-05$} & \revisiontext{$1.43e-24$}\\
\revisiontext{30\%} & \revisiontext{$2.68e-04$} & \revisiontext{$1.63e-06$}\\
\revisiontext{40\%} & \revisiontext{$5.95e-06$} & \revisiontext{$1.84e-07$}\\
\revisiontext{50\%} & \revisiontext{$6.15e-04$} & \revisiontext{$2.27e-03$}\\
\revisiontext{60\%} & \revisiontext{$4.45e-03$} & \revisiontext{$5.17e-03$}\\
\bottomrule
\end{tabular}
\end{table}

Finally, we emphasize that our method can have another benefit, i.e., the reduced computational cost. Machine learning models are increasing in size, thus increased computational cost. To illustrate this benefit, we measure computational time. For selected experiments, we run the experiments in an otherwise idle computer system and record wall-clock time. The computer system is equipped with an AMD EPYC 7742 64-Core Processor, 512 GB RAM, and a  TESLA V100S GPU. We compare to a baseline of the LineVul model trained with 100\% of its training data.

\begin{table}[tb]
\centering
\caption{Applying Patch Selection to Vulnerability Prediction (LineVul)}
\label{tab:rq2:linevul}
\begin{tabular}{r|r r|r r}
\toprule
Training Data & \multicolumn{2}{c|}{\bf Performance} & \multicolumn{2}{c}{\bf Computation Time} \\
   Selected~\% & F1 & Gain~\% & Time & Reduction \\
\midrule
\revisiontext{100\%} & \revisiontext{0.843} & ---- & \revisiontext{8h57m} & ----\\
\revisiontext{40\%}  & \revisiontext{0.872} & \revisiontext{$3.52\%$}  & \revisiontext{3h55m} & \revisiontext{$\frac{\text{8h57m} - \text{3h55m}}{\text{8h57m}} \approx 56\%$}     \\
\bottomrule
\end{tabular}
\end{table}

The experimental result in Table~\ref{tab:rq2:linevul} shows that the proposed patch curation algorithm, which selects 40\% of the training data, can benefit LineVul by gaining a small improvement in predictive performance of \revisiontext{3.52\% (Table~\ref{tab:linevul:hptest})}. However, at the same time, the training time is substantially reduced by approximately 56\%, i.e., a significant energy saving in training. \revisiontext{This benefit is particularly valuable for large vulnerability data collections (that may even include data augmentation), which are increasingly important to modern ML models.}

The observed improvement in predictive performance is limited by the
fact that the test partition extracted from the Big-Vul data used in the LineVul study is likely to contain some noise, as it was gathered directly using an automated approach based on a set of heuristics and has been reported to contain some labeling errors~\cite{ding2024vulnerability}. Furthermore, the evaluation is limited by the size of the dataset for LineVul, i.e., the Big-Vul dataset.

\begin{mdframed}
    \noindent
    \textit{
        A trained model of the proposed UQ-based data curation approach based on the EHAL heuristics can improve the accuracy and training time of software vulnerability prediction, a popular use of software vulnerability patch data. 
}
\end{mdframed} \section{Related Work}
\label{sec:related}

We divide the related work into two categories: 1) approaches for vulnerability patch curation; and 2) uncertainty quantification.

\subsection{Vulnerability Patch Identification}
The research community has been diligently contributing efforts to curate multiple vulnerability patch datasets, and a common method employed by these researchers is through manual reviews~\cite{rei_database_2017, lin_software_2020, jimenez_engineering_2018, wang_detecting_2019, fan_cc_2020, reis_ground-truth_2021, bhandari_cvefixes_2021, ponta_manually-curated_2019, nguyen_vulcurator_2022, vulncode-db}. More recently, researchers have started to scale vulnerability patch curation through machine learning, a semi-automated process where they train a machine learning model using a set of manually vetted vulnerability patches and subsequently identify vulnerability patches through inference~\cite{zhou_automated_2017, cabrera_commit2vec_2021, wang_patchrnn_2021, zhou_finding_2021, zhou_spi_2021, wu_enhancing_2022, tan_locating_2021, wang_vcmatch_2022}. Regarding vulnerability patch curation, we classify these efforts into two distinct categories, which we term as vulnerability patch prediction and vulnerability patch association. The former involves the classification of a changeset, whether it is one commit or a group of semantically related commits, into either a "general" vulnerability changeset (i.e., a changeset that fixes a vulnerability) or not~\cite{zhou_automated_2017, cabrera_commit2vec_2021, wang_patchrnn_2021, zhou_finding_2021, zhou_spi_2021, wu_enhancing_2022}. On the other hand, the latter aims to determine whether a changeset corresponds to a "specific type" of vulnerability changeset that addresses a known vulnerability, such as a CVE~\cite{tan_locating_2021, wang_vcmatch_2022}.

In this work, we make two essential contributions. First, we provide a principled approach to assess the quality of data curation~\cite{zhou_automated_2017, cabrera_commit2vec_2021, wang_patchrnn_2021, zhou_finding_2021, zhou_spi_2021, wu_enhancing_2022}. Second, we offer guidance on data curation, focusing on not only data quality but also data usefulness, through the application of Uncertainty Quantification (UQ) techniques~\cite{tan_locating_2021, wang_vcmatch_2022}. As such, our work is not simply to ``clean'' existing datasets, rather it is aimed to curate highly useful security patches for improving down-stream applications of the patches. \revisiontext{While dataset cleaning primarily addresses technical errors and inconsistencies, curation involves selecting data points that maximize utility, ensuring the resulting dataset is optimized for improving model performance and training efficiency.}

\subsection{Uncertainty Qualification}
In our work, we leverage recent advances in UQ for deep learning. Most UQ techniques are probabilistic approaches~\cite{abdar2021review, gawlikowski2021survey, mena2021survey, he2023survey}. We can contrast different UQ techniques along several dimensions, e.g., frequentist versus Bayesian approaches, single prediction versus set prediction approaches, direct modeling versus approximation~\cite{hullermeier2021aleatoric, gawlikowski2021survey}. Primarily, UQ has been leveraged to address ``AI safety'', for which, an essential problem is to understand to what extent the user can trust the decision made by a machine learning model~\cite{amodei_concrete_2016, gros_ai_2023}. Not until recently have researchers begun to use UQ to assess data quality for machine learning~\cite{seedat_data-iq_2022}. Our work here is not to propose a novel UQ approach for AI safety. Rather it is to investigate two interconnected issues~\cite{zhou_spi_2021,croft2023data}, the data quantity and the data quality of vulnerability patch data via UQ approaches that disentangle epistemic and aleatoric uncertainties~\cite{kendall_what_2017,depeweg2018decomposition}. Via evaluating combinations of data distribution modeling and UQ approximation techniques, our work paves a path for a principled approach to curate and select changesets for software security applications. In addition, our result show that the proposed approach can positively contribute to important research direction in security assurance, such as software vulnerability prediction.

 \section{Threats to Validity}
\label{sec:threat}

The primary threat to validity is the inability to generalize the findings, which map to internal threat to validity.

\noindent
{\bf Selection of UQ Models.}
There are a wealth of UQ research that have resulted in a myriad of UQ concepts and modeling approaches with varying assumptions, advantages, and pitfalls~\cite{he2023survey, hullermeier2021aleatoric}. Selecting a suitable UQ model is a challenging task for software vulnerability patch curation. This challenge is complicated by the lack of ground-truth uncertainty for real-world datasets. To evaluate ordinary machine learning tasks, such as regression and classification, we rely on ground truth values that are sometimes referred to as ``gold set'' data. For UQ studies on practical problems, there is a lack of UQ gold set data to evaluate UQ models. To address this threat, research commonly experiments with how UQ models respond to data quality shifts. Intuitively, when data quality shifts increases, we expect the models' prediction ability to degrade, which should results in lower predictive performance and higher uncertainty. In this approach, we experiment with multiple UQ modeling techniques to select the most suitable model for this study. 

\noindent
{\bf Generalization of Findings.}
Our research findings strongly suggest that Uncertainty Quantification (UQ) can serve as a systematic and principled approach to curate software changeset data, enabling effective vulnerability patch identification and other software quality assurance tasks. However, it is important to acknowledge a potential internal threat to the validity of these conclusions, as the generalizability of our results might be limited due to the number of models, datasets, and UQ approaches explored in our initial experiments. \revisiontext{Furthermore, we acknowledge that our technique has thus far been applied only to the SAP, VCMatch, and Big-Vul datasets.} To mitigate this internal validity concern, we conducted an extensive array of experiments. By exploring various combinations of UQ techniques, datasets, and data features, we sought to ensure robustness in our findings. The key conclusions consistently emerged across all of our experiments, reinforcing the reliability of our main research outcomes.

\ronerevisiontext{This study randomly splits labeled commit patches into training, validation, and test partitions, a standard practice in software vulnerability prediction research. While data leakage is a general concern in machine learning, random splits cannot fully prevent it due to potential dependencies between commits, such as change couplings and genealogies~\cite{d2009relationship,herzig2013predicting}. In our case, this risk is minimal. The datasets consist of vulnerability patches and randomly selected clean patches, without full evolutionary histories. As a result, the chance of leakage from commit dependencies is low.
}

\noindent {\revisiontext{\bf Sources of Uncertainty.}}
\revisiontext{
A recent study calls into the attention of the effectiveness of noisy label learning on deep learning for program understanding~\cite{wang2024empirical}. The study indicates that deep learning models still struggle to detect real-world noise in program understanding datasets. In this research, we do not differentiate the sources of uncertainty, i.e., whether they stem from the feature space or the labeling space. A future direction will be to investigate whether UQ can improve noise label learning for program understanding by more accurately detecting labeling errors. 
}

\noindent {\revisiontext{\bf Programming Languages and Data Selection.}}
\revisiontext{Programming Languages and Data Selection. In this paper, we evaluated the impact of the proposed data selection approach on software vulnerability prediction using C/C++ code. Whether this data selection approach has a similar impact on software vulnerability prediction for other programming languages or in a cross-language setting remains an open question for future exploration.}

 \section{Conclusion and Future Work}
\label{sec:summary}

Machine learning techniques that leverage historical vulnerabilities have become an important direction in software security assurance. However, obtaining high-quality vulnerability patch data, which is crucial for advancing this research, poses significant challenges. Several studies confirm substantial quality issues in software vulnerability patches collected from the NVD~\cite{tan_locating_2021,croft2023data}. Furthermore, there is a lack of a systematic approach to curate patch datasets and assess patch quality. Relying solely on software security experts for manual curation is impractical due to cost and limited availability of expertise.
In addition, prior works primarily focus on data quality alone~\cite{croft2023data}. As machine learning techniques become an important research direction software security assurance, whether curated patches are informative to machine learning should be an important dimension to examine. However, designing patch data curation approaches that account for both utility and quality is particularly challenging to design due to the fact that both quality and usefulness span a spectrum. 

To address these concerns and challenges, our work aims to develop an approach for curating software changeset data for security software assurance. We propose an automatic, machine learning-based security patch curation approach intended to select patches with high quality and utility value. This is made possible by developing a heuristic that leverages both epistemic and aleatoric uncertainties computed on security patches. From our evaluation, we observe the following: 1) Model Ensemble is an effective method for producing reliable Uncertainty Quantification (UQ) in security patch curation; 2) software changeset data can exhibit different distributions across projects or components, making \htsc modeling a superior approach that demonstrates better predictive performance and UQ quality compared to \hmsc modeling; 3) epistemic uncertainty can serve as a proxy for the informativeness of the patches while aleatoric uncertainty can serve as a guidance for the quality of the patches; \revisiontext{4) in all evaluations, we compare data curated using the proposed approach with 100\% of the data from VCMatch and Big-Vul, representing heuristic-based and machine learning-based state-of-the-art curation methods, respectively, and find that it selects higher-quality and more informative patches.} Our study highlights that UQ measures can guide the selection of changesets to benefit downstream secure software quality assurance tasks, such as a software vulnerability prediction model. For a manually curated high-quality dataset, the benefits come 1) a small but statistically significant improvement of predictive performance, and 2) a significant reduction in software vulnerability model training time (i.e., significant energy saving). Our work is a first to curate security patches with a focus on their utility values in addition to their quality, which sets the stage for the future work that aims to scale our experiments to other, larger vulnerability datasets and curate vulnerability patches from databases such as the NVD to significantly improve predictive performance and reduce uncertainty for machine learning-based software security assurance approaches. 

\revisiontext{Future work also includes assessing new UQ models aimed at enabling a data curation pipeline that further improves downstream applications.}
\ronerevisiontext{
Additionally, advances in large language models (LLMs) have led to the exploration of non-training methods, such as in-context learning and prompt engineering, in numerous application settings~\cite{wies2023learnability,brown2020language}. A recent study demonstrates the potential of generative LLMs in filtering vulnerability data by identifying and removing low-quality instances using carefully designed prompts~\cite{dil2025}. We posit that our work can also be used to curate high-utility, representative examples for prompts or in-context learning inputs, thereby enhancing the effectiveness of these non-training methods.
}

\balance
\bibliographystyle{IEEEtran}
\bibliography{vuldata_2022,VulnerabilityPrediction,weblinks,uq_chen,ai,vuldata,se}

\end{document}